%% file: FIM_ISAC.tex
\documentclass[lettersize,journal]{IEEEtran}
\usepackage{amsmath,amssymb,amsfonts}
\usepackage[utf8]{inputenc}
\usepackage{algorithmic}
\usepackage{algorithm}
\usepackage{array}
\usepackage[caption=false,font=scriptsize,labelfont=sf,textfont=sf]{subfig}
\usepackage{textcomp}
\usepackage{stfloats}
\usepackage{url}
\usepackage{verbatim}
\usepackage{graphicx}
\usepackage{cite}
\usepackage{amssymb}
\usepackage{setspace}
\usepackage{lineno} 
\usepackage{cases}
\usepackage{subfloat}
\usepackage{float}  
\usepackage{titlesec}
\usepackage{textcomp}
\usepackage{bm,bbold}
\usepackage{xcolor}
\usepackage{siunitx}
\input{sym_marco.tex}

\begin{document}

\title{Cram\'er-Rao Bound Minimization for Flexible Intelligent Metasurface-Enabled ISAC Systems}

\author{Qian Zhang,
		Yufei Zhao,
		Jiancheng An,~\IEEEmembership{Member,~IEEE,}
		Zheng Dong,~\IEEEmembership{Member,~IEEE,}\\
		Yong Liang Guan,~\IEEEmembership{Senior Member,~IEEE,}
		Ju Liu,~\IEEEmembership{Senior Member,~IEEE,}
		and 
		Chau Yuen,~\IEEEmembership{Fellow,~IEEE}
\thanks{
	The corresponding authors: Ju Liu; Zheng Dong. (e-mail: \{juliu, zhengdong\}@sdu.edu.cn.)
}
\thanks{Qian Zhang is with School of Information Science and Engineering, Shandong University, Qingdao 266237, China, and with School of Electrical and Electronic Engineering, Nanyang Technological University, Singapore (e-mail: qianzhang2021@mail.sdu.edu.cn).}
\thanks{Zheng Dong and Ju Liu are with School of Information Science and Engineering, Shandong University, Qingdao 266237, China (e-mail: zhengdong@sdu.edu.cn; juliu@sdu.edu.cn).}
\thanks{Yufei Zhao, Jiancheng An, Yong Liang Guan, and Chau Yuen are with School of Electrical and Electronics Engineering, Nanyang Technological University, Singapore 639798 (e-mail: yufei.zhao@ntu.edu.sg; jiancheng\_an@163.com; EYLGuan@ntu.edu.sg; chau.yuen@ntu.edu.sg).}

}

\maketitle

\begin{abstract}
Integrated sensing and communication (ISAC) have been widely recognized as a key enabler for future wireless networks, where the Cram\'er-Rao bound (CRB) plays a central role in quantifying sensing accuracy.
In this paper, we present the first study on CRB minimization in flexible intelligent metasurface (FIM)-enabled ISAC systems.
Specifically, we first derive an average CRB expression that explicitly depends on FIM surface shape and demonstrate that array reconfigurability can substantially reduce the CRB, thereby significantly enhancing sensing performance.
Moreover, to tackle the challenging CRB minimization problem, we adopt average Fisher information maximization as a surrogate objective and use the Gauss-Hermite quadrature method to obtain an explicit approximation of the objective function.
The resulting problem is then decoupled into three subproblem, i.e., beamforming optimization and transmit/receive FIM surface shape optimization.
For beamforming optimization, we employ the Schur complement and penalty-based semi-definite relaxation (SDR) technique to solve it.
Furthermore, we propose a fixed-point equation method and a projected gradient algorithm to optimize the surface shapes of the receive and transmit FIMs, respectively.
Simulation results demonstrate that, compared to rigid arrays, surface shaping of both transmit and receive FIMs can significantly reduce the average sensing CRB while maintaining communication quality, and remains effective even in multi-target scenarios.

\end{abstract}

\vspace{-0.1cm}

\vspace{-0.4cm}
\section{Introduction} 
With the rapid evolution of wireless systems, an increasing number of applications demand not only reliable communication but also high-precision sensing, such as smart cities and autonomous driving~\cite{Liu2022ISAC,Zhang2025FA_ISAC}.
As a result, integrated sensing and communication (ISAC) has garnered significant attention from both academia and industry, and extensive research efforts have been devoted to its development under various scenarios~\cite{Zhang2025FA_ISAC,Liu2022CRB,Zhang2025SIM_ISAC,Liu2024RIS_ISAC}.
For instance, in~\cite{Liu2022CRB}, Liu {\it et al.} proposed a beamforming optimization approach for ISAC systems that minimizes the sensing Cram\'er-Rao bound (CRB) while guaranteeing communication quality of services (QoS), thereby enhancing parameter estimation accuracy.
In~\cite{Zhang2025SIM_ISAC}, Zhang {\it et al.} investigated the trade-off between sensing beam gain and communication rates in stacked intelligent metasurfaces (SIM)-assisted ISAC systems to improve target detection performance.
In~\cite{Liu2024RIS_ISAC}, Liu {\it et al.} examined target detection and parameter estimation in reconfigurable intelligent surfaces (RIS)-assisted ISAC systems to enhance sensing capability.
Most of these studies rely on conventional rigid arrays (RA), where the transmit beamformer is designed to balance communication and sensing.
However, RA cannot reconfigure the array surface geometry, leading to limited degrees of freedom (DoF), which in turn constrains the ISAC performance and limits their ability to support more demanding applications.

Fortunately, the concept of a flexible intelligent metasurface (FIM) has recently been introduced as a reconfigurable surface shape technology~\cite{Bai2022FIM_Nature,Ni2022FIM_NC,An2025FIM_TAP,An2025FIM,An2025FIM_TCOM}.
FIM enables an additional degree of structural flexibility beyond conventional RA by allowing rapid reconfiguration of its surface shapes.
For example, while a conventional uniform linear array is typically arranged along the $x$-axis, FIM technology permits radiating elements displacement in the orthogonal $y$-axis, thereby providing dynamic spatial adaptability~\cite{An2025FIM,An2025FIM_TCOM}.
This technological advancement is particularly promising, as it breaks through the limitations of traditional ISAC designs and offers extra DoF to further enhance system performance.
In~\cite{Teng2025FIM_ISAC}, Teng {\it et al.} proposed an FIM-enhanced multi-target sensing framework, where joint optimization of FIM surface shape and beamformer significantly improved target detection capability.
In~\cite{Ranasinghe2025FIM_ISAC}, Ranasinghe {\it et al.} investigated high-mobility ISAC systems and demonstrated that introducing FIM not only preserves sensing capability but also enhances communication performance.
These studies have shown that FIM can effectively reduce the false alarm probability and thereby improves target detection accuracy.
However, whether FIM can also enhance the parameter estimation capability in ISAC systems remains an open question.

Building upon the above analysis, we present the first study on average CRB minimization in FIM-enabled ISAC systems.
Based on the sensing prior distribution, we derive the average CRB expression, which explicitly demonstrates that FIM surface shaping can significantly reduce the CRB and thereby enhance sensing performance.
To tacking the challenging QoS-constrained CRB minimization problem, we first employ Bayesian average information maximization as a surrogate for the original objective, then convert the objective function into a closed-form expression using the Gauss-Hermite quadrature method.
Furthermore, we propose a beamforming optimization framework that combines the Schur complement with a penalty-based semi-definite relaxation (SDR) algorithm. 
By leveraging dual theory, we derive a fixed-point equation to obtain the optimal surface shape of the receive FIM. 
Finally, we develop a projected gradient ascent (PGA) algorithm to optimize the transmit FIM surface shape and employ the increasing penalty dual decomposition (IPDD) algorithm to decouple complex constrained projection subproblems for efficient solution.
Simulation results verify that the effective FIM surface shaping achieves lower CRB than traditional RA.

{\it{Notation:}} 
$(\cdot)^{*}$, $(\cdot)^{\rm T}$ and $(\cdot)^{\rm H}$ denote the conjugate, transpose, and conjugate transpose operation, matrix inverse operation, respectively.
$\mathbb{C}$ and $\mathbb{R}$ represent complex and real space, respectively.
${\cal CN}(\cdot,\cdot)$ denotes the circularly symmetric complex Gaussian distribution.
${\cal R}\{\cdot\}$ and ${\cal I}\{\cdot\}$ denote the real and imaginary parts, respectively.
${\rm vec}(\cdot)$ and ${\rm Tr}\left\{\cdot \right\}$ represent the vectorization and trace of the matrix, respectively.
$j$ represents the imaginary unit.
$\bm 1_N$ is a full-one vector with $N$ elements.
$\bold{I}_K$ denotes a $K \times K$ identity matrix.
$[\bz]_n$ denotes the $n$-th entry of the vector $\bz$.
$\left\| \cdot \right\|_{*}$, $\|\cdot\|_2$, and $|\cdot|$ denote the unclear norm, Euclidean norm, and absolute value, respectively.
$\Pi_{{\cal C}}$ denotes a projection operation.

\begin{figure}[t]	
	\centering \includegraphics[width=0.75\linewidth]{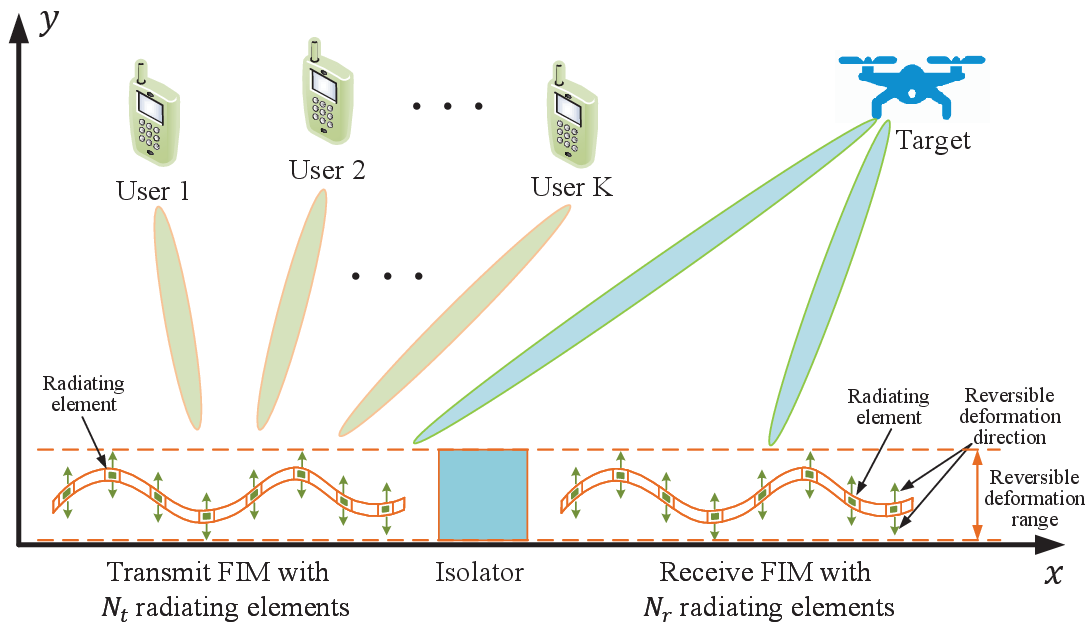}
	\vspace{-0.3cm}
	\caption{The model for FIM-enabled ISAC systems.}
	\vspace{-0.6cm}
	\label{fig:System_Model}
\end{figure}

\vspace{-0.35cm}
\section{System Model} \vspace{-0.1cm}
As shown in Fig.~\ref{fig:System_Model}, we investigate an FIM-enabled ISAC system, 
where the transmitter employs a linear FIM array to enable effective communications and sensing, and the receiver equipped with a linear FIM to enhance parameter estimation accuracy.
In such a system, the base station (BS), equipped with a transmit FIM comprising $N_t$ radiating elements and a receive FIM with $N_r$ radiating elements, simultaneously supports communication with $K$ single-antenna users and sensing of a target.
In addition, we require $N_r\geq N_t$ to avoid information loss in target sensing.
In the ISAC framework, the achievable rate of users and the CRB are adopted to characterize the communication and sensing performance, respectively, with detailed mathematical models provided in the following subsections.

\vspace{-0.45cm}
\subsection{Signal Model} 
Let $\bX \in \mathbb{C}^{N_t \times T}$ be an ISAC signal matrix, with $T>N_t$ being the length of the radar pulse/communication frame.
The matrix $\bX$ is given by \vspace{-0.15cm}
\begin{equation}
	\begin{split}
		\bX = \bW_C \bS_C + \bW_R \bS_R = \left[ \bx[1], \bx[2],\dots, \bx[T] \right]^{\rm T},
	\end{split}
\end{equation}
\vspace{-0.55cm}

\noindent where $\bW_C \in \mathbb{C}^{N_t\times K}$ is the dual-functional beamforming matrix for both communications and sensing. 
$\bS_C\in \mathbb{C}^{K\times T}$ contains $K$ unit-power data streams intended for the $K$ users.
$\bW_R \in \mathbb{C}^{N_t\times N_r}$ is the beamforming matrix for sensing and $\bS_R \in \mathbb{C}^{N_r \times T}$ is the dedicated sensing signal~\cite{Liu2022CRB,Liu2024RIS_ISAC}.

For convenience, we define $\bS = \left[ \bS_C^{\rm T}~~ \bS_R^{\rm T} \right]^{\rm T}$ and $\bW = \left[ \bW_C ~~ \bW_R \right] = \left[ \bw_1,\dots,\bw_{K+N_r} \right]\in \mathbb{C}^{N_t\times (K+N_r)} $.
In addition, we assume that $\frac{1}{T} \bS \bS^{\rm H} \approx\bI_{K+N_r}$ when the block length $T$ is sufficiently large~\cite{Liu2022CRB}.

Therefore, at the $t$-th slot, the received signal at user $k$ is  \vspace{-0.15cm}
\begin{equation}
	\begin{split}
		y_k[t] = \bh_k^{\rm H} \bx[t] + n_k[t],
	\end{split}
\end{equation}
\vspace{-0.55cm}

\noindent where $n_k[t] \sim {\cal CN}\left( 0, \sigma_k^2 \right)$ denotes the noise at user $k$, with $\sigma_{k}^{2}$ representing the noise power.
The channel $\bh_k\in \mathbb{C}^{N_t\times 1}$ with respect to user $k$ can be represented as \vspace{-0.15cm}
\begin{equation}
	\begin{split}
		\bh_k = \sum_{\ell=1}^{L_k} \alpha_k^\ell \ba(\by^t,\theta_k^\ell),
	\end{split}
\end{equation}

\vspace{-0.3cm}\noindent
 where $\alpha_k^\ell \in \mathbb{C}$ denotes the path loss of the $\ell$-th path with respect to user $k$, $\theta_k^\ell$ denotes the angle of the $\ell$-th path with respect to user $k$, and $L_k$ denotes the number of paths.

By transmitting $\bX$ to sense a target, the received echo signal reflected by the target at the receiver of the BS is given by \footnote{
This work considers high-frequency regimes (e.g., 30 GHz), where line-of-sight (LOS) propagation dominates and non-LOS multi-path components are relatively weak. Such weak multi-path or Gaussian-distributed clutter can be modeled as an effective increase in the noise power, which does not alter the analytical structure of the CRB but only modifies the associated weighting terms. Consequently, the proposed CRB-oriented optimization framework remains directly applicable in the presence of clutter.
} \vspace{-0.15cm}
\begin{equation}
	\begin{split}
		\bY_R = \alpha_r \bb(\by^r,\theta_t) \ba^{\rm H}(\by^t,\theta_t)\bX + \bN_R,
	\end{split}
\end{equation}
\vspace{-0.6cm}

\noindent where $\bN_R \in \mathbb{C}^{N_r\times T}$ denotes the additive white Gaussian noise (AWGN) matrix at the receiver, with variance of each entry being $\sigma_r^2$, $\alpha_r \in \mathbb{C}$ denotes the reflection coefficient that depends on both the radar cross section (RCS) and the round-trip path loss, 
$\theta_t$ denotes the angle of the target, which is random parameter following a prior Gaussian distribution ${\cal N}(\overline{\theta},\sigma_\theta^2)$.
$\bb(\by^r,\theta_t)$ and $\ba(\by^t,\theta_t)$ denote respectively the receiver- and transmitter-target steering vectors as follows \vspace{-0.1cm}
\begin{equation*}
	\begin{split}
		\left[ \ba(\by^t,\theta_t) \right]_n &= e^{j\delta \left( x_{n}^t \cos\theta_t + y_{n}^t  \sin\theta_t \right)},~n=1,2,\dots,N_t, \\
		\left[\bb(\by^r,\theta_t) \right]_n &= e^{j\delta \left( x_{n}^r \cos\theta_t + y_{n}^r  \sin\theta_t \right)},~n=1,2,\dots,N_r,
	\end{split}
\end{equation*}
\vspace{-0.25cm}

\noindent 
where $\delta = \frac{2\pi}{\lambda}$, $\lambda$ is the wavelength.
$\bx^t=\left[ x_1^t,\dots,x_{N_t}^t \right]^{\rm T}$ and $\bx^r =\left[ x_1^r,\dots,x_{N_r}^r \right]^{\rm T}$ represent the coordinates of the radiating elements at the transmitter and receiver, respectively.
Specifically, 
$x_n^t = d_x\times (n-1),1\leq n \leq N_t$ and $x_n^r = d_x\times (n-1),~1\leq n \leq N_r$, where $d_x$ denotes the inter-element spacing. 
	$\by^t=\left[y_1^t,\dots,y_{N_t}^t \right]^{\rm T}$ and
$\by^r=\left[y_1^r,\dots,y_{N_r}^r \right]^{\rm T}$ characterize the surface shape of the FIM, which allows adjustment within a reversible deformation range to satisfy \vspace{-0.3cm}
$$
{\cal C}_{y^r}: y_n^r \in [y_{\rm min}, y_{\rm max}],~n=1,2,\dots,N_r,
$$
$$
{\cal C}_{y^t}: y_n^t \in [y_{\rm min}, y_{\rm max}],~n=1,2,\dots,N_t.
$$

\vspace{-1cm}
\subsection{Performance Metrics for Communications} 
In such an ISAC system, we expect to ensure that each user can obtain an achievable rate that is not less than the threshold rate $R_{\rm th}$.
Therefore, for communication, the following inequality must be satisfied. \vspace{-0.15cm}
\begin{equation*}
	\begin{split}
		{\cal C}_k: \gamma_k \triangleq \frac{ | \bh_k^{\rm H} \bw_k  |^2}{\sum_{i=1,i\neq k}^{K+N_r}| \bh_k^{\rm H} \bw_i  |^2 + \sigma_k^2   } \geq r_{\rm th},~k=1,2,\dots,K,
	\end{split}
\end{equation*}
\vspace{-0.4cm}

\noindent where $\gamma_k$ denotes the signal-to-interference-plus-noise ratio (SINR) of user $k$, $r_{\rm th} = 2^{R_{\rm th}}-1$, and $R_{\rm th}$ denotes the achievable rate threshold.

\vspace{-0.35cm}
\subsection{Performance Metrics for Sensing} \vspace{-0.05cm}
We derive the average CRB to characterize the lower bound for the unbiased sensing parameter estimation error.
Specifically, we define the unknown parameter vector to be estimated as $\bm\xi = \left[\theta_t, {\cal R}\{\alpha_r\}, {\cal I}\{\alpha_r\}  \right]^{\rm T}$.

We first vectorize the received signal $\bY_R$ as \vspace{-0.15cm}
\begin{equation}
	\begin{split}
		\by_R = \alpha_r {\rm vec}\left( \bb(\by^r,\theta_t) \ba^{\rm H}(\theta_t)\bW \bS \right) + \bm n_R,
	\end{split}
\end{equation}
\vspace{-0.5cm}

\noindent where $\bm n_R \triangleq {\rm vec}(\bN_R)$.

Define $\bA = \bb(\by_r,\theta_t) \ba^{\rm H}(\theta_t)$, the $(i,m)$-th element of the Fisher information matrix $\bF_{\rm IM} \in \mathbb{C}^{3\times 3} $ can be obtained as~\cite{Kay1993SSP}
\begin{equation*} \label{Fisher_matrix}
	\begin{split}
		\left[ \bF_{\rm IM} \right]_{i,m} = \frac{2}{\sigma_r^2} {\cal R}\left\{ \frac{\partial \alpha_r {\rm vec}^{\rm H}\left( \bA \bW \bS \right)}{\partial \xi_i}  \frac{\partial \alpha_r {\rm vec}\left( \bA \bW \bS \right)}{\partial \xi_m} \right\}.
	\end{split}
\end{equation*}

We can derive each entry of $\bF_{\rm IM}$ as \vspace{-0.1cm}
\begin{equation} \label{Fisher_information_matrix}
	\begin{split}
		F_{\theta\theta} &= \frac{2T|\alpha_r|^2}{\sigma_r^2} {\rm Tr}\left\{ \dot{\bA} \bW \bW^{\rm H} \dot{\bA}^{\rm H} \right\}, \\
		\bF_{\theta\bm\alpha} &= \frac{2T}{\sigma_r^2} {\cal R}\left\{ {\rm Tr}\left\{ \alpha_r^* \bA \bW \bW^{\rm H} \dot{\bA}^{\rm H} \right\} [1~j] \right\}, \\
		\bF_{\bm\alpha\bm\alpha} &= \frac{2T}{\sigma_r^2} {\rm Tr}\left\{ \bA \bW \bW^{\rm H} \bA^{\rm H} \right\} \bI_2, 
	\end{split}
\end{equation}
\vspace{-0.35cm}

\noindent where $\dot{\bA}$ represents the derivative of $\bA$ with respect to $\theta_t$.

In order to improve sensing performance with the aid of FIM, we aim to further derive the explicit expression of Fisher information. 
Specifically, we can obtain \vspace{-0.15cm}
\begin{equation*}
	\begin{split}
		{\rm Tr}\left\{ \bA \bW \bW^{\rm H} \bA^{\rm H} \right\} = {\rm Tr}\left\{ \bb^{\rm H} \bb \ba^{\rm H} \bW \bW^{\rm H} \ba \right\} = N_r \left\| \ba^{\rm H} \bW \right\|_2^2,
	\end{split}
\end{equation*}
\begin{equation*}
	\begin{split}
		{\rm Tr}\left\{ \bA \bW \bW^{\rm H} \dot{\bA}^{\rm H} \right\} =&  j\delta\left\| \ba^{\rm H} \bW \right\|_2^2  \bm 1_{N_r}^{\rm T} \bm\zeta + N_r \ba^{\rm H} \bW \bW^{\rm H} \dot{\ba}, \\
		{\rm Tr}\left\{ \dot{\bA} \bW \bW^{\rm H} \dot{\bA}^{\rm H} \right\} =& \delta^2 \left\| \ba^{\rm H} \bW \right\|_2^2 \bm\zeta^{\rm T} \bm\zeta + N_r \left\| \dot{\ba}^{\rm H} \bW \right\|_2^2 \\
		&+ 2 \delta \cdot {\cal I}\left\{ \ba^{\rm H} \bW \bW^{\rm H} \dot{\ba} \right\} \bm 1_{N_r}^{\rm T} \bm\zeta,
	\end{split}
\end{equation*}
\vspace{-0.2cm}

\noindent where $\dot{\ba} = \left[\dot{a}_1,\dots,\dot{\ba}_{N_t} \right]^{\rm T}$, $\dot{a}_n = -j \delta x_n \sin\theta_t e^{j\delta x_n \cdot \cos\theta_t }$, and $\bm\zeta(\theta_t) = \sin\theta_t \bx^r - \cos\theta_t \by^r$.

We derive the Fisher information with respect to $\theta_t$ as
\begin{equation*} \label{Fisher_information}
	\begin{split}
		{\rm F}(\theta_t)= &\frac{2T |\alpha_r|^2}{\sigma_r^2}\left(\bm\zeta^{\rm T} \bP \bm\zeta \left\| \ba^{\rm H} \bW \right\|_2^2  + N_r \left\| \dot{\ba}^{\rm H} \bW \right\|_2^2 \right) \\
		&-\frac{2T |\alpha_r|^2}{\sigma_r^2} \cdot \frac{N_r \left| \ba^{\rm H} \bW\bW^{\rm H} \dot{\ba} \right|^2 }{\left\| \ba^{\rm H} \bW \right\|_2^2},
	\end{split}
\end{equation*}
where $\bP = \delta^2\bI_{N_r} - \frac{\delta^2}{N_r} \bm 1_{N_r} \bm 1_{N_r}^{\rm T}$.

Therefore, the average CRB with respect to $\theta_t$ can be formulated as $\overline{{\rm C}}_{\theta_t } = \mathbb{E}_{\theta_t}\left[ 1/{\rm F}(\theta_t) \right]$.

\vspace{-0.2cm}
\section{Cram\'er-Rao Bound Minimization} 
To enhance ISAC performance, we aim to reduce CRB while satisfying QoS requirements of users by jointly designing the beamformer and the FIM surface shape.
However, due to the non-explicit nature of the expectation over inverse operations for CRB, as well as the strong coupling among different parameter realizations, it is difficult to directly minimize the average CRB by adjusting the beamformer and the surface shape.
To address this issue, we adopt a tractable surrogate objective by maximizing the average Fisher information, which satisfies $\overline{{\rm C}}_{\theta_t } \geq 1/ \mathbb{E}_{\theta_t}\left[ {\rm F}(\theta_t) \right]$.
From an information-theoretic perspective, maximizing $\mathbb{E}_{\theta_t}\left[ {\rm F}(\theta_t) \right]$ enhances the overall information content collected over the parameter uncertainty region, thereby enabling a robust sensing-oriented design~\cite{Li2010ACRB,Aharon2024ECRB}.
Under prior distribution $\theta_t \sim {\cal N}(\overline{\theta},\sigma_\theta^2)$, this metric differs from the Bayesian Fisher information only by a constant term~\cite{Liu2025BCRB,Aharon2024ECRB}.
Therefore, we can reformulate the optimization problem as \footnote{This paper primarily focuses on single-target sensing to validate the advantages of FIM in enhancing ISAC performance. For multi-target scenarios, the objective function can be naturally extended to the sum of average Fisher information across multiple targets, i.e, $\sum_{i}\mathbb{E}_{\theta_t^i}\left[ {\rm F}(\theta_t^i) \right]$. In this case, the proposed algorithm remains effective in solving the corresponding problem. } \vspace{-0.2cm}
\begin{equation} \label{CRB_QoS}
	\begin{split}
		\max_{ \bW,\by^r,\by^t } \mathbb{E}_{\theta_t}\left[ {\rm F}(\theta_t) \right] ~
		\mbox{s.t.}	\,
		{\cal C}_p: \left\| \bW \right\|_{\rm F}^2 \leq P_{\rm max}, {\cal C}_{y^r}, {\cal C}_{y^t},{\cal C}_k.
	\end{split}
\end{equation}
\vspace{-0.5cm}

However, the objective function in problem~\eqref{CRB_QoS} is not available in closed form. 
Therefore, we seek a closed-form surrogate objective by approximating the expectation integral with a finite summation~\cite{Aharon2024ECRB,Shapiro2009CRB}.
Specifically, we have
\begin{equation*}
	\begin{split}
		\mathbb{E}_{\theta_t}\left[ {\rm F}(\theta_t) \right] 
		= \frac{1}{\sqrt{\pi}} \int_{-\infty}^{+\infty} {\rm F}(\overline{\theta} + \sqrt{2}\sigma_\theta z) e^{-t^2} dz 
		 \overset{\text{(a)}}{\approx} \sum_{u=1}^{U} \omega_u {\rm F}(\tilde\theta_u),
	\end{split}
\end{equation*}
where $\tilde\theta_u = \overline{\theta} + \sqrt{2}\sigma_\theta z_u$ denotes the $u$-th sample point, $\omega_u$ is the corresponding weighted value, and the step $\text{(a)}$ is derived by Gauss-Hermite quadrature method~\cite{Gradshteyn2007Integrals}.

Therefore, we can reformulate the problem as \vspace{-0.2cm}
\begin{equation} \label{FI_QoS}
	\begin{split}
		&\max_{ \bW,\by^r,\by^t }~ \sum_{u=1}^{U} \omega_u {\rm F}(\tilde\theta_u) \\
		&\mbox{s.t.}	~
		{\cal C}_p, ~{\cal C}_{y^r},~ {\cal C}_{y^t},~ {\cal C}_k,\, k=1,\dots,K.
	\end{split}
\end{equation}

\vspace{-0.15cm}
The above problem is highly non-convex due to complicated constraints and coupled variables.
In order to solve it, we first decouple optimization variables by the alternating optimization (AO) algorithm.


\subsection{Optimize $\bW$ Given $\by^r$ and $\by^t$}
Given $\by^r$ and $\by^t$, the problem~\eqref{FI_QoS} with respect to $\bW$ is given by  \vspace{-0.1cm}
\begin{equation} \label{CRB_W}
	\begin{split}
		\max_{ \bW } ~ \sum_{u=1}^{U} \omega_u {\cal G}_u(\bW) ~~
		\mbox{s.t.}	
		~ {\cal C}_p,~ {\cal C}_k,\, k=1,\dots,K,
	\end{split}
\end{equation}
where $\kappa_u = \bm\zeta^{\rm T}(\tilde\theta_u) \bP \bm\zeta(\tilde\theta_u)$, $\ba_u = \ba(\by^t,\tilde\theta_u)$, and 
$
{\cal G}_u(\bW) =  \kappa_u \| \ba_u^{\rm H} \bW \|_2^2 + N_r \left\| \dot{\ba}_u^{\rm H} \bW \right\|_2^2 - N_r \frac{\left| \ba_u^{\rm H} \bW\bW^{\rm H} \dot{\ba}_u \right|^2}{\left\| \ba_u^{\rm H} \bW \right\|_2^2}.
$

The objective function of the problem~\eqref{CRB_W} is a complicated fractional structure. 
By introducing auxiliary variable $\{\gamma_1,\dots,\gamma_U \}$, $\bE_k = \bw_k \bw_k^{\rm H}$, and $\widetilde\bE = \sum_{k=1}^{K+N_r}\bE_k$, we use the Schur complement method~\cite{Zhang2005Schur} and the SDR algorithm~\cite{luo2010sdr} to transform it into the following problem.
\begin{equation} \label{CRB_W_Shur}
	\begin{split}
		&\max_{ \bW, \{\gamma_u\}_{u=1}^{U} } ~\sum_{u=1}^{U} \omega_u \left\{ \gamma_u \cdot N_r + \kappa_u \cdot {\rm Tr}\left\{ \ba_u\ba_u^{\rm H} \widetilde\bE  \right\} \right\}   \\
		&\mbox{s.t.}	
		~ {\rm Tr}\left\{\widetilde\bE \right\} \leq P_{\rm max},~ \left\| \bE_k \right\|_{*} - \left\| \bE_k \right\|_{2} = 0,  \\
		& \left[
		\begin{array}{cc}
			{\rm Tr}\left\{ \dot{\ba}_u\dot{\ba}_u^{\rm H}\widetilde{\bE} \right\} - \gamma_u & {\rm Tr}\left\{ \dot{\ba}_u \ba_u^{\rm H} \widetilde{\bE} \right\} \\
			{\rm Tr}\left\{ \ba_u \dot{\ba}_u^{\rm H} \widetilde{\bE}  \right\} & {\rm Tr}\left\{ \ba_u \ba_u^{\rm H} \widetilde{\bE} \right\}
		\end{array}
		\right] \succeq 0, \\
		&~ \frac{{\rm Tr}\left\{ \bh_i \bh_i^{\rm H}\bE_i \right\}}{r_{\rm th}} - \sum_{k=1,k\neq i}^{K+N_r} {\rm Tr}\left\{ \bh_i\bh_i^{\rm H} \bE_k \right\} \geq \sigma_i^2,
	\end{split}
\end{equation}
where $\left\| \bE_k \right\|_{*} - \left\| \bE_k \right\|_{2} = 0$ ensures rank-one property of $\bE_k$.

To guarantee the rank-one property of solutions, we treat constraint $\left\| \bE_k \right\|_{*} - \left\| \bE_k \right\|_{2} = 0$ as a penalty function.
In addition, we have $\left\| \bE_k \right\|_{*} - \left\| \bE_k \right\|_{2} > 0$ when the rank of $\bE_k$ is more than one.
Therefore, the objective function can be transformed into
\begin{equation*}
	\begin{split}
		\sum_{u=1}^{U} \omega_u \left\{ \gamma_u \cdot N_r + \kappa_u \cdot {\rm Tr}\left\{ \ba_u\ba_u^{\rm H} \widetilde\bE  \right\} \right\} - \frac{1}{\rho}\left(\left\| \bE_k \right\|_{*} - \left\| \bE_k \right\|_{2}\right),
	\end{split}
\end{equation*}
where $\rho\geq 0$ is a given penalty factor.

Next, we employ the successive convex approximation (SCA) algorithm to transform the above objective into the following concave function.
\begin{equation*}
	\begin{split}
		\sum_{u=1}^{U} \omega_u \left\{ \gamma_u \cdot N_r + \kappa_u \cdot {\rm Tr}\left\{ \ba_u\ba_u^{\rm H} \widetilde\bE  \right\} \right\} - \frac{1}{\rho}\left(\left\| \bE_k \right\|_{*} + \tilde\bE_k^n \right),
	\end{split}
\end{equation*}
where $\tilde\bE_k^n = -\left\|\bE_k^n \right\|_2 - {\rm Tr}\left\{ \bv_{{\rm max},k}^n \bv_{{\rm max},k}^{n,{\rm H}} \left( \bE_k - \bE_k^n \right) \right\} $, $\bE_k^n$ denotes the solution at the $n$-th iteration, and $\bv_{{\rm max},k}^n$ denotes the eigenvector corresponding to the largest eigenvalue of $\bE_k^n$.
Therefore, the problem~\eqref{CRB_W_Shur} can be transformed into a convex problem and solved efficiently.

\vspace{-0.2cm}
\subsection{Optimize $\by^r$ Given $\bW$ and $\by^t$}
Given $\bW$ and $\by^t$, the problem~\eqref{FI_QoS} with respect to $\by^r$ can be reformulated as
\begin{equation} \label{CRB_zeta}
	\begin{split}
		\max_{ \by^r } ~ \by^{r,{\rm T}} \bQ \by^r - 2 \bq^{\rm T} \by^r ~~
		\mbox{s.t.}	
		~ {\cal C}_{y^r},
	\end{split}
\end{equation}
where $\bq = \tilde\eta_1\bP\bx^r$, $\tilde\eta_1 = \sum_{u=1}^{U} \omega_u \left\| \ba_u^{\rm H} \bW \right\|_2^2 \sin\tilde\theta_u \cos\tilde\theta_u $, $\bQ = \tilde\eta_2\bP$, and $\tilde\eta_2 = \sum_{u=1}^{U} \omega_u \left\| \ba_u^{\rm T}\bW \right\|_2^2 (\cos\tilde\theta_u)^2 $.

Noting that $\bQ \succeq 0$, the optimal solution to problem~\eqref{CRB_zeta} is obtained at the boundary of the feasible set.
To proceed, we introduce the indicator vector $\bm J = \left[J_1,J_2,\dots,J_{N_r} \right]^{\rm T}$ where $J_n \in \{0,1\}$ for $n=1,\dots,N_r$.
Furthermore, setting $d_y = y_{\rm max} - y_{\rm min}$, the optimal solution can be characterized by  \vspace{-0.15cm}
\[
y_n^r = y_{\rm min} + d_y \cdot J_n,~n=1,2,\dots,N_r.
\]

\vspace{-0.25cm}
Then, we can transform the problem~\eqref{CRB_zeta} into \vspace{-0.15cm}
\begin{equation} \label{zeta_J_1}
	\begin{split}
		\max_{ \bm J } &~ {\cal G}(\bm J) = \sum_{n=1}^{N_r} \beta_n \cdot J_n - \frac{1}{N_r}\left( \sum_{n=1}^{N_r} d_y \cdot J_n \right)^2  \\
		\mbox{s.t.}	
		&~ J_n \in \{0,1\},~n = 1,2,\dots,N_r,
	\end{split}
\end{equation}
where $ \beta_n = \tilde\eta_2 d_y^2 - 2\tilde\eta_1d_y x_n^r + 2\tilde\eta_2 d_y y_{\rm min} $.

According to
$
-\frac{S^2}{N} = \min_{\eta \in \mathbb{R}}~\left( N\eta^2 - 2\eta S \right),
$
we have \vspace{-0.1cm}
\begin{equation} 
	\begin{split}
		{\cal G}(\bm J) = \min_{\eta \in \mathbb{R}}~\left( N_r \eta^2 + \sum_{n=1}^{N_r} \left(\beta_n - 2\eta d_n \right)J_n \right).
	\end{split}
\end{equation}

\vspace{-0.1cm}
Based on the duality theory~\cite{Boyd2004convex}, we can obtain the following inequality. \vspace{-0.1cm}
\begin{equation}
	\begin{split}
		\max_{\bm J} ~{\cal G}(\bm J) = \max_{\bm J} \min_{\eta}~ {\cal L}(\bm J, \eta) \leq \min_{\eta}\max_{\bm J}~ {\cal L}(\bm J, \eta),
	\end{split}
\end{equation}
\vspace{-0.5cm}

\noindent where ${\cal L}(\bm J, \eta) = N_r \eta^2 + \sum_{n=1}^{N_r} \left(\beta_n - 2\eta d_y \right)J_n$.

Strong duality holds if and only if there exists an $(\bm J^\star, \eta^\star)$ such that for all feasible $(\bm J, \eta)$. In this case, we have~\cite{Boyd2004convex} \vspace{-0.1cm}
\begin{equation}
	\begin{split}
		{\cal L}(\bm J^\star, \eta) \leq {\cal L}(\bm J^\star, \eta^\star) \leq {\cal L}(\bm J, \eta^\star).
	\end{split}
\end{equation}

\vspace{-0.2cm}
Therefore, we aim to find feasible solution to satisfy the above condition.

{\it 1) Given $\eta$:}
The optimal $J_n$ is given by \vspace{-0.1cm}
\begin{equation} 
	J_n = \arg\min_{J_n\in\{0,1\}}~{\cal L}(\bm J, \eta^\star)= 
	\begin{cases}
		1,~{\rm if}~ \tau_n \geq \eta^\star, \\
		0,~{\rm if}~ \tau_n < \eta^\star,
	\end{cases}
\end{equation}
\vspace{-0.2cm}

\noindent where $\tau_n = \frac{\beta_n}{2d_y}$.



{\it 2) Given $\bm J$:}
The optimal $\eta$ is given by \vspace{-0.1cm}
\begin{equation} 
	\eta = \arg\min_{\eta\in \mathbb{R}}~{\cal L}(\bm J^\star, \eta)= \frac{\sum_{n=1}^{N_r} d_y\cdot J_n^\star}{N_r}.
\end{equation}

%

\vspace{-0.1cm}
In summary, saddle point $(\bm J^\star, \eta^\star)$ must satisfy the following fixed-point equation as \vspace{-0.1cm}
\begin{equation}
	\begin{split}
		\eta^\star = \frac{d_y}{N_r} \sum_{n: \tau_n \geq \eta^\star} 1.
	\end{split}
\end{equation}

\vspace{-0.1cm}
For the discrete case, we can reformulate the fixed-point equation as $\eta_m = \frac{md_y}{N_r}$.
We then arrange $\tau_n$ in ascending order, i.e.
$
\tau_1 \leq \tau_2\leq \dots \leq \tau_{N_r}.
$
The fixed point condition is equivalent to the existence of a $m$ satisfying
$
\tau_m \leq \eta_m \leq \tau_{m+1}.
$
Therefore, the optimal $J_n$ is selected as 0 when $1\leq n \leq m$ and $1$ when $m+1 \leq n \leq N_r$.
Otherwise, the optimal value is obtained at $m=1$ or $m=N_r$.
We then can obtain $m = \lceil N_r\tilde\eta_1/2 + N_r(\tilde\eta_2y_{\rm min} - \tilde\eta_1 x_n^r)/d_y \rceil$.

\vspace{-0.3cm}
\subsection{Optimize $\by^t$ Given $\bW$ and $\by^r$}
Given $\bW$ and $\by^r$, the problem~\eqref{FI_QoS} with respect to $\by^t$ can be reformulated as
\begin{equation} \label{CRB_yt}
	\begin{split}
		\max_{ \by^t } ~ {\cal F}(\by^t) ~~
		\mbox{s.t.}	
		~ {\cal C}_{y^t},~{\cal C}_k,\, k=1,2,\dots,K,
	\end{split}
\end{equation}
where ${\cal F}(\by^t) = \sum_{u=1}^{U} \omega_u {\cal G}_u(\by^t)$.

The problem~\eqref{CRB_yt} is highly non-convex since the objective and constraints ${\cal C}_k$ are non-convex.
To deal with the problem~\eqref{CRB_yt}, we propose the PGA algorithm, which follows
\begin{equation*}
	\begin{split}
		(\tilde\by^t)^{i+1} = (\by^t)^{i} + \tilde\rho \nabla_{\by^t}{\cal F}(\by^t),~(\by^t)^{i+1} = \Pi_{{\cal C}_k \cap {\cal C}_{y^t} }\left((\tilde\by^t)^{i+1} \right),
	\end{split}
\end{equation*}
where $\tilde\rho > 0$ is a given ascent step length, $\nabla_{\by^t}{\cal F}(\by^t)$ denotes the gradient of ${\cal F}(\by^t)$ with respect to $\by^t$. We then have
\begin{equation*}
	\begin{split}
		&\frac{\partial {\cal G}_u(\by^t)}{\partial y_n^t} = 2 \kappa_u \Re\left\{ \bg_u^{n,{\rm H}} \widetilde\bE \ba_u \right\} + 2 N_r\Re\left\{ \hat\bg_u^{n,{\rm H}} \widetilde\bE \dot{\ba}_u \right\} - \\
		&2N_r \left( \frac{ \Re\left\{\overline{\bg}^* \left[\bg_u^{n,{\rm H}} \widetilde\bE \dot{\ba}_u + \ba_u^{\rm H} \widetilde{\bE}\hat\bg_u^n \right] \right\} }{ \Re\left\{ \ba_u^{\rm H} \widetilde{\bE} \ba_u \right\}  } - \frac{\left|\overline{\bg} \right|^2\Re\left\{ \bg_u^{n,{\rm H}} \widetilde\bE \ba_u \right\}     }{ \left(\Re\left\{ \ba_u^{\rm H} \widetilde{\bE} \ba_u \right\} \right)^2 } \right),
	\end{split}
\end{equation*}
where $\hat\bg_u^n = \left[j\delta \sin\tilde\theta_u(\ba_u + \dot{\ba}_u) \right] \odot \be_n$, $\bg_u^n = \left[j\delta \sin\tilde\theta_u\ba_u \right] \odot \be_n$, $\overline{\bg} = \ba_u^{\rm H}\widetilde{\bE} \dot{\ba}_u $, $\widetilde{\bE} = \bW\bW^{\rm H}$, and $\be_n \in \mathbb{R}^{N_t\times 1}$ is a vector where only the $n$-th element is 1 and all others are 0.

Further, the projection $\Pi_{{\cal C}_{k} \cap {\cal C}_{y^t}} (\bm\xi)$ can be obtained by solving the following problem
\begin{equation} \label{proj_yt}
	\begin{split}
		\min_{\by^t}~\| \by^t - \bm\xi \|_2^2 ~~ {\rm s.t.}~ {\cal C}_{y^t},~{\cal C}_k,\, k=1,2,\dots,K.
	\end{split}
\end{equation}
\vspace{-0.5cm}

Then, we employ the IPDD algorithm to solve the problem~\eqref{proj_yt}, which follows the updates \vspace{-0.1cm}
\begin{subequations}
	\begin{align}
		(\by^t)^{i+1} = &~\mathop{\rm argmin}\limits_{\by^t \in {\cal C}_{y^t}} \| \by^t - \bm\xi \|_2^2 + {\cal H}(\by^t, \bh_k^{i},\bm\upsilon_k^i),\label{IPDD_step_1}\tag{21a}
	\end{align}
\end{subequations}
\vspace{-0.55cm}
\begin{subequations}\label{IPDD_step}
	\begin{align}
		\bh_k^{i+1} = &~\mathop{\rm argmin}\limits_{\bh_k \in {\cal C}_{k}} ~ {\cal H}((\by^t)^{i+1}, \bh_k,\bm\upsilon_k^i),  \label{IPDD_step_2} \tag{21b} \\
		\bm\upsilon_k^{i+1} =&~\bm\upsilon_k^{i} + \left(\bh_k^{i+1} - \sum_{\ell=1}^{L_k} \alpha_k^\ell \ba((\by^t)^{i+1},\theta_k^\ell) \right) \big/ \mu, \label{IPDD_step_3} \tag{21c}
	\end{align}
\end{subequations} 

\vspace{-0.25cm}\noindent
for $k=1,2,\dots,K$, where $\mu>0$ is a given penalty factor and ${\cal H}(\by^t,\bh_k,\bm\upsilon_k) =  \frac{1}{2\mu} \sum_{k=1}^{K} \| \bh_k - \sum_{\ell=1}^{L_k} \alpha_k^\ell \ba(\by^t,\theta_k^\ell) + \mu \bm\upsilon_k \|_2^2$.

\vspace{-0.3cm}
For the subproblem~\eqref{IPDD_step_1}, we can also use the projected gradient algorithm to solve it.
For the subproblem~\eqref{IPDD_step_2}, the objective function is convex, but the constraints are non-convex.
To handle it, we use the SCA algorithm to transform the constraints into the convex form
$
2\Re\left\{ \hat\bh_k^{\rm H} \bw_k \bw_k^{\rm H} \bh_k \right\} - \left| \hat\bh_k^{\rm H} \bw_k \right|^2 \geq r_{\rm th}\left( \sum_{i=1,i\neq k}^{K+N_r}| \bh_k^{\rm H} \bw_i  |^2 + \sigma_k^2  \right),
$
for $k=1,2,\dots,K$, where $\hat\bh_k$ is the solution from the previous iteration.
Therefore, the subproblem~\eqref{IPDD_step_2} can be transformed into a convex problem and solved effectively.
On the other hand, the optimal solution of the problem~\eqref{CRB_yt} can also be obtained by line-search-based element-wise method~\cite{Zhang2024PRIS}.

\begin{figure}[t]	
	\centering \includegraphics[width=0.58\linewidth]{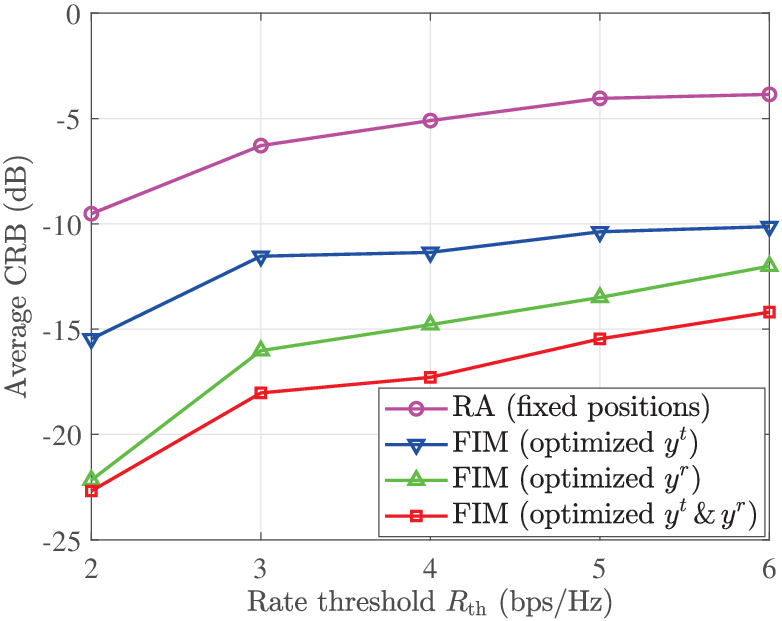}
	\vspace{-0.3cm}
	\caption{Average CRB values under different achievable rate threshold $R_{\rm th}$ with morphing range $\lambda$.}
	\vspace{-0.4cm}
	\label{fig:CRB_R_th}
\end{figure}
\begin{figure}[t]	
	\centering \includegraphics[width=0.58\linewidth]{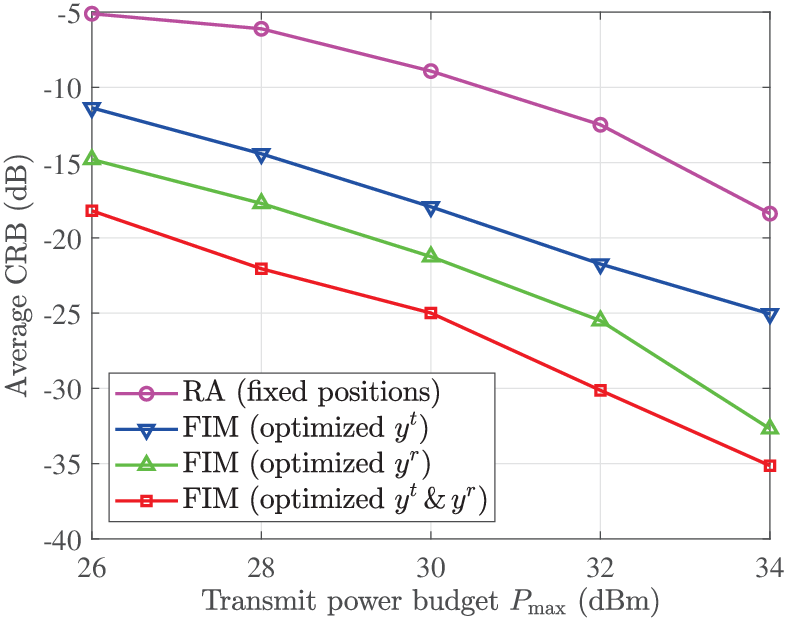}
	\vspace{-0.3cm}
	\caption{Average CRB values under different transmit power budgets $P_{\rm max}$ with morphing range $\lambda$, $R_{\rm th}=4\,$bps/Hz.}
	\vspace{-0.45cm}
	\label{fig:CRB_P_max}
\end{figure}
\begin{figure}[t]	
	\centering \includegraphics[width=0.58\linewidth]{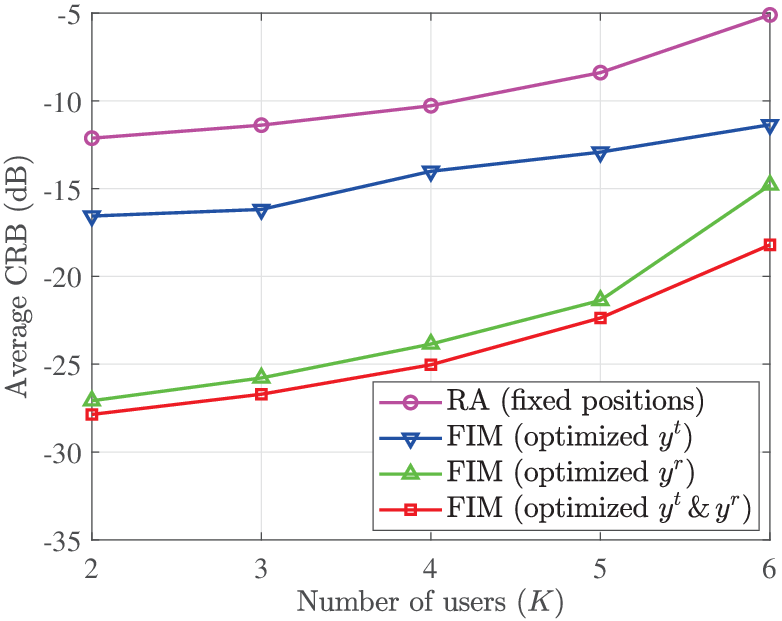}
	\vspace{-0.3cm}
	\caption{Average CRB values under different number of users with morphing range $\lambda$, $R_{\rm th}=4\,$bps/Hz.}
	\vspace{-0.5cm}
	\label{fig:CRB_K}
\end{figure}
\begin{figure}[t]	
	\centering \includegraphics[width=0.58\linewidth]{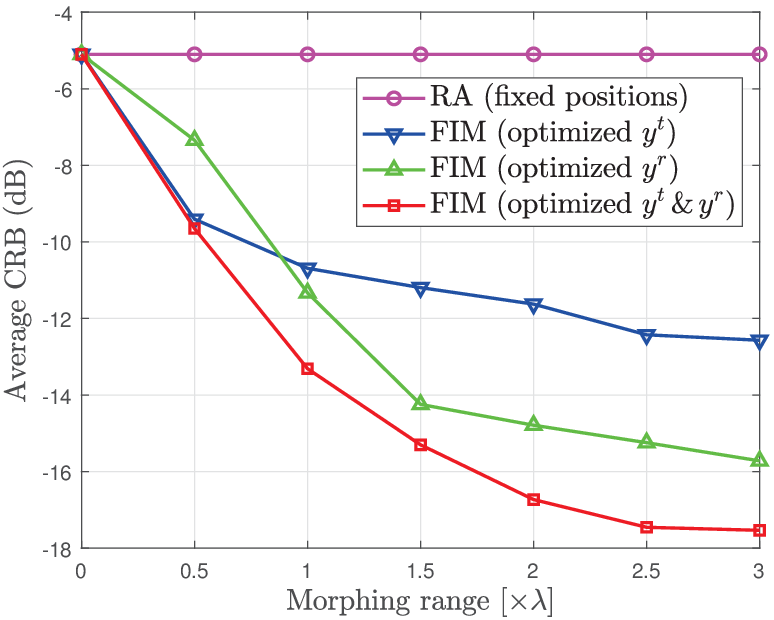}
	\vspace{-0.3cm}
	\caption{Average CRB values obtained when the morphing range is between $0$ and $3\lambda$, $R_{\rm th}=4\,$bps/Hz.}
	\vspace{-0.65cm}
	\label{fig:CRB_kappa}
\end{figure}
\begin{figure}[t]
	\centering
	\vspace{-0.35cm}
	\subfloat[Beampattern ]{\includegraphics[width=0.5\columnwidth]{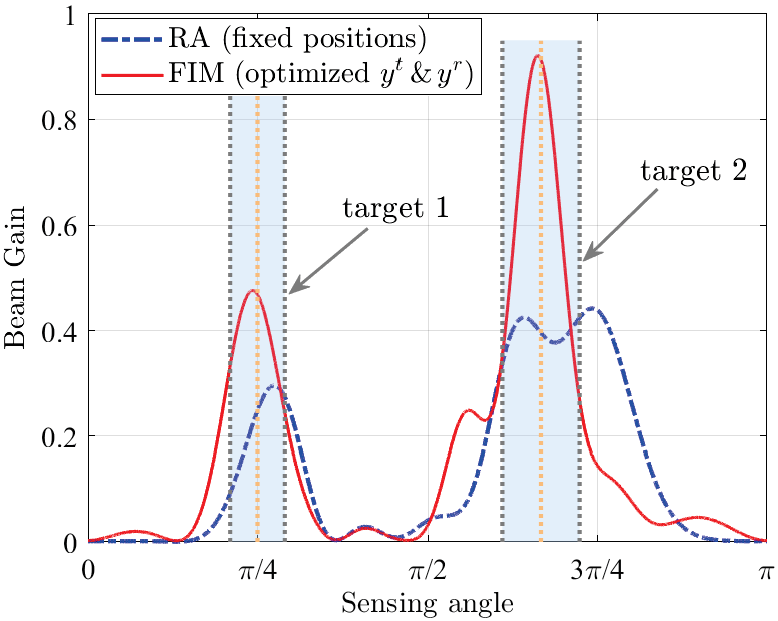}%
		}
	\hfil
	\subfloat[MUSIC spectrum ]{\includegraphics[width=0.5\columnwidth]{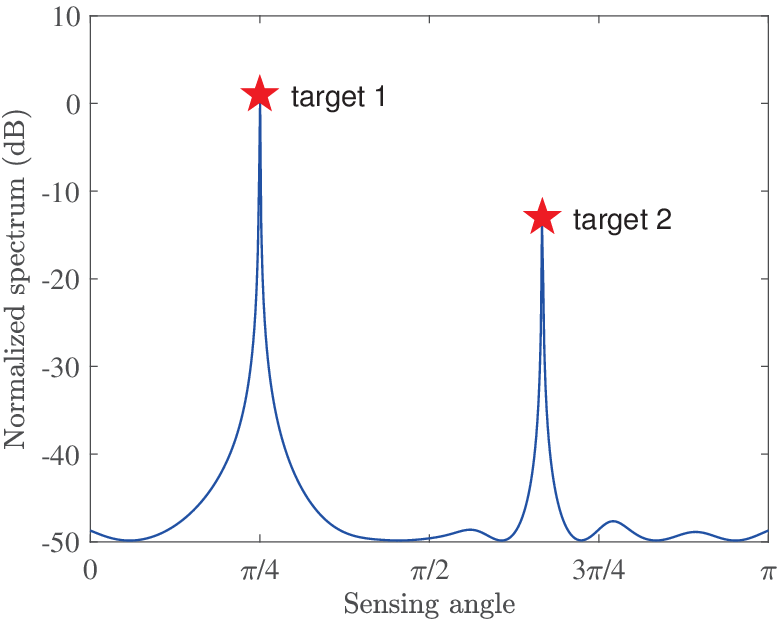}%
		}
	\vspace{-0.05cm}
	\caption{Beampattern obtained through RA and FIM for a two-target scenario and corresponding MUSIC spectrum, $R_{\rm th}=4\,$bps/Hz.}
	\vspace{-0.5cm}
	\label{fig:Beam_pattern}
\end{figure}

\vspace{-0.2cm}
\section{Simulation Results}\vspace{-0.1cm}
In this section, we present numerical results to validate the effectiveness of the proposed algorithm and to illustrate the CRB improvement enabled by the FIM.
For the simulation setup, we consider $\lambda=0.01\,$m, $N_t=N_r=8$, $d_x=\lambda/2$, $y_{\rm min}=0$, $y_{\rm max}=2\lambda$, $T = 128$, $\overline{\theta}=\pi/4$, $\sigma_\theta = \frac{0.4\times0.886}{N_t \cos\overline{\theta} }$, $L_k=1$, $P_{\rm max}=26\,$dBm, and a noise power of -80$\,$dBm.
In the objective function of the problem~\eqref{FI_QoS}, we set $\omega_u = \frac{2^{U-1} U! \sqrt{\pi} }{U^2 \left[H_{U-1}(z_u) \right]^2}$ and $U=5$, where $H_{U}(\cdot)$ denotes the $U$-th order Hermite polynomial function, and $z_u$ is the $u$-th root of fifth order Hermite polynomial function.
Furthermore, the path loss of BS-user channels are modeled as $\alpha_k^\ell = 10^{-3}d_k^{-2.2}$, where $d_k$ denotes the distance between the transmitter and user $k$.
The users are randomly and uniformly located within a distance range of 30 to 80 meters from BS.

Fig.~\ref{fig:CRB_R_th} illustrates the average CRB performance under different achievable rate thresholds $R_{\rm th}$. 
It can be observed that, as $R_{\rm th}$ increases, the average CRB of all schemes exhibits an increasing trend. 
This is because more stringent QoS constraints force a larger portion of resources to be allocated to communication users, which inevitably sacrifices sensing performance. 
Nevertheless, compared with the RA-based scheme, the proposed FIM-based surface shape optimization scheme can significantly reduce the average CRB, i.e., improve the sensing performance, under various communication rate thresholds. 
In particular, optimizing only the transmit FIM surface shape achieves a lower CRB than optimizing only the receive FIM surface shape, since the transmit FIM needs to jointly support both communication and sensing, whereas the receive FIM is dedicated solely to sensing. 
Moreover, the scheme with joint optimization of the transmit and receive FIM surface shapes consistently delivers the best performance, indicating a clear complementarity between the transmit and receive array geometries in terms of Fisher information gain. 
These results demonstrate that, even under relatively high communication rate constraints, the proposed FIM surface shape optimization method can effectively mitigate the degradation of sensing performance.

It can be observed from Fig.~\ref{fig:CRB_P_max} that, as $P_{\rm max}$ increases, the average CRB of all schemes decreases significantly, since higher transmit power improves the echo signal-to-noise ratio, thereby directly enhancing the Fisher information. 
Moreover, the FIM-based scheme consistently outperforms the RA-based scheme.
Fig.~\ref{fig:CRB_K} illustrates that, as the increase of $K$, multi-user interference becomes more severe and the competition for resources intensifies, leading to an increased CRB for all schemes. 
Furthermore, the FIM-based scheme still significantly outperforms the RA-based scheme, which achieves the minimum CRB across different user scales. 
Therefore, by properly adjusting the FIM surface, the spatial resolution and parameter estimation accuracy can be effectively improved.

From Fig.~\ref{fig:CRB_kappa}, it can be observed that, as the increase of morphing range, the CRB of the FIM-based schemes continuously decreases, whereas the performance of the RA-based scheme remains almost unchanged. 
This result indicates that the surface reconfigurability is one of the key factors for improving sensing performance. 
A larger geometric flexibility enables the array to better match the target direction distribution, thereby increasing the effective aperture and spatial resolution. 
Meanwhile, the joint optimization of the transmit and receive FIM surface shapes again exhibits the best performance, validating the important role of cooperative reconfiguration of the transmit and receive arrays in CRB minimization.

Moreover, we extend the proposed method to multi-target scenarios. 
Fig.~\ref{fig:Beam_pattern}$\,$(a) compares the beam patterns of the RA-based and FIM-based schemes in a two-target scenario. 
Since the prior distribution of the sensing directions is known, the system design aims to align the beam toward a sensing region rather than a single precise angle. 
It can be observed that, compared with the beam generated by the RA-based scheme, optimizing the FIM surface shape enables stronger beam energy concentration within the sensing region. 
From a spatial-domain perspective, this phenomenon provides an intuitive explanation for the aforementioned CRB performance improvement: through FIM surface shaping, the system can focus more energy toward the sensing directions, thereby effectively improving the condition number and the effective information content of the Fisher information matrix, and ultimately achieving higher-accuracy target parameter estimation.
As shown in~\ref{fig:Beam_pattern}$\,$(b), based on the optimized beamformer and FIM surface shape, we can use the multiple signal classification (MUSIC) method to accurately estimate the targets' angle.

\section{Conclusion}
In this paper, we presented the first study on average CRB minimization in FIM-enabled ISAC systems by explicitly incorporating sensing prior distributions. 
Although the average CRB generally does not admit a closed-form expression, we showed that effective FIM surface shaping can significantly improve sensing performance. 
To handle the resulting QoS-constrained optimization problem, Bayesian average information maximization was adopted as a surrogate objective and efficiently approximated via Gauss–Hermite quadrature method. A structured solution framework was then developed, including penalty-based SDR for beamformer optimization, a fixed-point equation method for receive FIM surface shaping, and a PGA algorithm combined with IPDD algorithm for transmit FIM surface shaping. 
Simulation results demonstrated that the proposed joint transmit–receive FIM surface shaping approach consistently achieves substantially lower average CRB than conventional RA-based designs, validating its effectiveness and robustness in ISAC systems.


%
%
%
%
%
%

\bibliographystyle{IEEEtran}
\bibliography{refs}

\end{document}

%% file: sym_marco.tex
\newcommand\bw{\ensuremath{{\bm w}}}
\newcommand\bW{\ensuremath{{\bold W}}}
\newcommand\bN{\ensuremath{{\bold N}}}

\newcommand\bq{\ensuremath{{\bm q}}}
\newcommand\bx{\ensuremath{{\bm x}}}
\newcommand\by{\ensuremath{{\bm y}}}

\newcommand\bh{\ensuremath{{\bm h}}}

\newcommand\be{\ensuremath{{\bm e}}}
\newcommand\bz{\ensuremath{{\bm z}}}

\newcommand\bP{\ensuremath{{\bold P}}}

\newcommand\bX{\ensuremath{{\bold X}}}

\newcommand\ba{\ensuremath{{\bm a}}}

\newcommand\bA{\ensuremath{{\bold A}}}

\newcommand\bb{\ensuremath{{\bm b}}}

\newcommand\bg{\ensuremath{{\bm g}}}

\newcommand\bF{\ensuremath{{\bold F}}}

\newcommand\bv{\ensuremath{{\bm v}}}

\newcommand\bY{\ensuremath{{\bold Y}}}

\newcommand\bS{\ensuremath{{\bold S}}}
\newcommand\bE{\ensuremath{{\bold E}}}

\newcommand\bQ{\ensuremath{{\bold Q}}}

\newcommand{\bI}{{\bold I}}

%% file: FIM_ISAC.bbl
\begin{thebibliography}{10}
\providecommand{\url}[1]{#1}
\csname url@samestyle\endcsname
\providecommand{\newblock}{\relax}
\providecommand{\bibinfo}[2]{#2}
\providecommand{\BIBentrySTDinterwordspacing}{\spaceskip=0pt\relax}
\providecommand{\BIBentryALTinterwordstretchfactor}{4}
\providecommand{\BIBentryALTinterwordspacing}{\spaceskip=\fontdimen2\font plus
\BIBentryALTinterwordstretchfactor\fontdimen3\font minus
  \fontdimen4\font\relax}
\providecommand{\BIBforeignlanguage}[2]{{%
\expandafter\ifx\csname l@#1\endcsname\relax
\typeout{** WARNING: IEEEtran.bst: No hyphenation pattern has been}%
\typeout{** loaded for the language `#1'. Using the pattern for}%
\typeout{** the default language instead.}%
\else
\language=\csname l@#1\endcsname
\fi
#2}}
\providecommand{\BIBdecl}{\relax}
\BIBdecl

\bibitem{Liu2022ISAC}
{A. Liu {\it et al.}}, ``A survey on fundamental limits of integrated sensing
  and communication,'' \emph{IEEE Commun. Surv. Tuts.}, vol.~24, no.~2, pp.
  994--1034, 2022.

\bibitem{Zhang2025FA_ISAC}
Q.~Zhang, M.~Shao, T.~Zhang, G.~Chen, J.~Liu, and P.~C. Ching, ``An efficient
  sum-rate maximization algorithm for fluid antenna-assisted {ISAC} system,''
  \emph{IEEE Commun. Lett.}, vol.~29, no.~1, pp. 200--204, Jan. 2025.

\bibitem{Liu2022CRB}
F.~Liu, Y.~F. Liu, A.~Li, C.~Masouros, and Y.~C. Eldar, ``Cramér-rao bound
  optimization for joint radar-communication beamforming,'' \emph{IEEE Trans.
  Signal Process.}, vol.~70, pp. 240--253, 2022.

\bibitem{Zhang2025SIM_ISAC}
Q.~Zhang, Z.~Du, Y.~Zhao, Y.~L. Guan, J.~Liu, and C.~Yuen, ``Joint power
  allocation and discrete phase-shift optimization for {SIM}-aided {ISAC}
  systems,'' \emph{IEEE Trans. Veh. Technol.}, pp. 1--6, 2025.

\bibitem{Liu2024RIS_ISAC}
R.~Liu, M.~Li, Q.~Liu, and A.~L. Swindlehurst, ``{SNR/CRB}-constrained joint
  beamforming and reflection designs for {RIS-ISAC} systems,'' \emph{IEEE
  Trans. Wireless Commun.}, vol.~23, no.~7, pp. 7456--7470, Jul. 2024.

\bibitem{Bai2022FIM_Nature}
{Y. Bai, H. Wang, Y. Xue, Y. Pan, J.-T. Kim, X. Ni, T.-L. Liu, Y. Yang, M. Han,
  Y. Huang {\it et al.}}, ``A dynamically reprogrammable surface with
  self-evolving shape morphing,'' \emph{Nature}, vol. 609, no. 7928, pp.
  701--708, Sept. 2022.

\bibitem{Ni2022FIM_NC}
{X. Ni, H. Luan, J.-T. Kim, S. I. Rogge, Y. Bai, J. W. Kwak, S. Liu, D. S.
  Yang, S. Li, S. Li {\it et al.}}, ``Soft shape-programmable surfaces by fast
  electromagnetic actuation of liquid metal networks,'' \emph{Nature Commun.},
  vol.~13, no.~1, p. 5576, Sept. 2022.

\bibitem{An2025FIM_TAP}
J.~An, M.~Debbah, T.~J. Cui, Z.~N. Chen, and C.~Yuen, ``Emerging technologies
  in intelligent metasurfaces: Shaping the future of wireless communications,''
  \emph{IEEE Trans. Antennas Propag.}, pp. 1--1, 2025.

\bibitem{An2025FIM}
J.~An, C.~Yuen, M.~D. Renzo, M.~Debbah, H.~V. Poor, and L.~Hanzo, ``Flexible
  intelligent metasurfaces for downlink multiuser {MISO} communications,''
  \emph{IEEE Trans. Wireless Commun.}, vol.~24, no.~4, pp. 2940--2955, Apr.
  2025.

\bibitem{An2025FIM_TCOM}
J.~An, Z.~Han, D.~Niyato, M.~Debbah, C.~Yuen, and L.~Hanzo, ``Flexible
  intelligent metasurfaces for enhancing {MIMO} communications,'' \emph{IEEE
  Trans. Commun.}, pp. 1--1, 2025.

\bibitem{Teng2025FIM_ISAC}
Z.~Teng, J.~An, L.~Gan, N.~Al-Dhahir, and Z.~Han, ``Flexible intelligent
  metasurface for enhancing multi-target wireless sensing,'' \emph{IEEE Trans.
  Veh. Technol.}, pp. 1--6, 2025.

\bibitem{Ranasinghe2025FIM_ISAC}
K.~R.~R. Ranasinghe, J.~An, I.~A.~M. Sandoval, H.~S. Rou, G.~T.~F. de~Abreu,
  C.~Yuen, and M.~Debbah, ``Flexible intelligent metasurfaces in high-mobility
  {MIMO} integrated sensing and communications,'' \emph{arXiv preprint
  arXiv:2507.18793}, 2025.

\bibitem{Kay1993SSP}
S.~M. Kay, \emph{Fundamentals of Statistical Signal Processing: Estimation
  Theory}.\hskip 1em plus 0.5em minus 0.4em\relax Upper Saddle River, NJ, USA:
  Prentice-Hall, 1993.

\bibitem{Li2010ACRB}
Y.~Li, H.~Minn, and J.~Zeng, ``An average cramer-rao bound for frequency offset
  estimation in frequency-selective fading channels,'' \emph{IEEE Trans.
  Wireless Commun.}, vol.~9, no.~3, pp. 871--875, Mar. 2010.

\bibitem{Aharon2024ECRB}
O.~Aharon and J.~Tabrikian, ``Asymptotically tight bayesian cramér-rao
  bound,'' \emph{IEEE Trans. Signal Process.}, vol.~72, pp. 3333--3346, 2024.

\bibitem{Liu2025BCRB}
R.~Liu, M.~Li, and A.~L. Swindlehurst, ``Joint array partitioning and
  beamforming designs in {ISAC} systems: A bayesian {CRB} perspective,''
  \emph{IEEE J. Sel. Areas Commun.}, 2025.

\bibitem{Shapiro2009CRB}
A.~Shapiro, D.~Dentcheva, and A.~Ruszczynski, \emph{Lectures on stochastic
  programming: Modeling and theory}.\hskip 1em plus 0.5em minus 0.4em\relax
  Philadelphia, PA: SIAM, 2009.

\bibitem{Gradshteyn2007Integrals}
I.~Gradshteyn and I.~Ryzhik, \emph{Table of integrals, series, and
  products}.\hskip 1em plus 0.5em minus 0.4em\relax 7th ed. New York, NY, USA:
  Academic, 2007.

\bibitem{Zhang2005Schur}
F.~Zhang, \emph{The Schur Complement and Its Applications}, vol. 4,Berlin,
  Germany: Springer-Verlag, 2005.

\bibitem{luo2010sdr}
Z.~Q. Luo, W.~K. Ma, A.~M.~C. So, Y.~Ye, and S.~Zhang, ``Semidefinite
  relaxation of quadratic optimization problems,'' \emph{IEEE Signal Process.
  Mag.}, vol.~27, no.~3, pp. 20--34, 2010.

\bibitem{Boyd2004convex}
S.~Boyd, S.~P. Boyd, and L.~Vandenberghe, \emph{Convex Optimization}.\hskip 1em
  plus 0.5em minus 0.4em\relax Cambridge, U.K.: Cambridge Univ. Press, 2004.

\bibitem{Zhang2024PRIS}
Q.~Zhang, J.~Liu, H.~Tang, Z.~Dong, and Y.~Li, ``Practical {RIS}-aided
  multiuser communications with imperfect {CSI}: Practical model, amplitude
  feedback, and beamforming optimization,'' \emph{IEEE Trans. Wireless
  Commun.}, vol.~23, no.~10, pp. 15\,245--15\,260, Oct. 2024.

\end{thebibliography}
